\documentclass[A4paper,12pt]{article}

\usepackage{fancyhdr}
\usepackage{anysize}

\marginsize{1.5cm}{1.5cm}{1cm}{4cm}

\usepackage{graphicx}
\usepackage{amsmath}
\usepackage{amsfonts}
\usepackage{epsfig}
\usepackage{lineno}
\usepackage{color}

\date{}
\begin{document}
\title{Classification of Boundary Gravitons in AdS$_3$ Gravity}

\author{Alan Garbarz$^{\dag,*}$, Mauricio Leston$^{\ddag,**}$}

\maketitle

\vspace{.2cm}

\begin{minipage}{.9\textwidth}\small \it \begin{center}
    $^\dag$ Instituto de F\'isica de La Plata (IFLP), CONICET $\&$ Departamento de F\'isica, Universidad Nacional de la Plata,
C.C. 67, (1900) La Plata, Argentina.
 \\  \end{center}
\end{minipage}

\vspace{.2cm}

\begin{minipage}{.9\textwidth}\small \it \begin{center}
    $^\ddag$ Instituto de Astronom\'ia y F\'isica del Espacio, Pabell\'on IAFE-CONICET, Ciudad Universitaria, C.C. 67 Suc. 28, Buenos Aires
     \end{center}
\end{minipage}

\vspace{.5cm}

\begin{abstract}

We revisit the description of the  space of asymptotically AdS$_3$ solutions of pure gravity in three dimensions with a negative cosmological constant as a collection of coadjoint orbits of the Virasoro group. Each orbit corresponds to a set of metrics related by diffeomorphisms which do not approach the identity fast enough at the boundary. Orbits contain more than a single element and this fact manifests the global degrees of freedom of AdS$_3$ gravity, being each element of an orbit what we call boundary graviton. 
We show how this setup allows to learn features about the classical phase space that otherwise would be quite difficult. Most important are the  proof of energy bounds and the characterization of boundary gravitons unrelated to BTZs and AdS$_3$. 
In addition, it makes manifest the underlying mathematical structure of the space of solutions close to infinity. Notably, because of the existence of a symplectic form in each orbit, being this related with the usual Dirac bracket of the asymptotic charges, this approach is a natural starting point for the quantization of different sectors of AdS3 gravity. We finally discuss previous attempts to quantize coadjoint orbits of the Virasoro group and how this is relevant for the formulation of  AdS$_3$ quantum gravity. 

\end{abstract}

\vspace{3cm}

\begin{flushleft}
\footnotesize
\parbox{\textwidth}{\mbox{}\hrulefill\\[-4pt]}
        {\scriptsize$^{*}$ E-mail: alan@df.uba.ar\\
\scriptsize$^{**}$ E-mail: mauricio@iafe.uba.ar}
\end{flushleft}       

\newpage

\section{Introduction}

In \cite{MaloneyWitten}, Maloney and Witten aimed to compute the partition function of AdS$_3$ gravity. To achieve this, the first step was to consider the space of solutions close to AdS$_3$ spacetime - its \textit{boundary gravitons} - namely all the metrics related to AdS$_3$ by diffeomorphisms that do not approach the identity fast enough at infinity. This gives an underlying Virasoro symmetry \cite{BrownHenneaux}  which when combined with previous results, allowed  the authors of \cite{MaloneyWitten} to arrive to an expression of the  partition function of this sector. The second step in \cite{MaloneyWitten} was to relate the AdS$_3$ sector just mentioned with the sectors of Euclidean black holes  through modular transformations and sum all the contributions of distinct sectors to  get the full partition function. Unfortunately, the final result obtained in this way was shown not to be physically sensible since  it cannot be written as the trace over a Hilbert space of the exponential of  a Hermitian operator (the Hamiltonian plus the angular momentum operator).

Let us focus on the computation of the partition function on the AdS$_3$ sector. As we mentioned above, this is the space of asymptotic geometries connected with AdS$_3$ by a non-trivial diffeomorphism. This particular partition function was obtained with the aid of previous results  on the quantization of coadjoints orbits of Virasoro group \cite{Witten88}. The quantization of coadjoint orbits was first introduced by Kirillov in order to study unitary representations of non-compact Lie groups (see for example \cite{Kirillov}). The reason why this becomes important for three-dimensional gravity with a negative cosmological constant is the following: The space of solutions, fulfilling Brown-Henneaux boundary conditions \cite{BrownHenneaux},  is naturally organized by coadjoint orbits of Virasoro group. Each orbit contains all the asymptotic geometries that are related by non-trivial diffeomorphisms. For example, the orbit of AdS$_3$ contains all the boundary gravitons of this geometry, namely all the non-trivial asymptotic diffeomorphisms of AdS$_3$. In \cite{Castro}, Castro et.  	al. implicitly used the coadjoint orbit structure of (some of) the boundary gravitons of AdS$_3$ to gain some insight into the spectrum of the quantum theory.  The same coadjoint orbit structure exists for the BTZ black holes \cite{BTZ}, and this was used in \cite{Nakatsu,Navarro99} with the intention to reproduce the entropy of the black holes. In addition, coadjoint orbits are manifolds endowed with a natural symplectic structure which, in the case of Virasoro group, is related to the usual Dirac bracket of the Brown-Henneaux charges. This fact makes relevant the description of the phase space of AdS$_3$ gravity as a collection of coadjoint orbits.

It would be desirable to have a quantization of  AdS$_3$ gravity, namely a quantization of the whole collection of coadjoint orbits of Virasoro group, but this seems far from reach. As a first step, we can learn a lot from the quantization of each orbit separately.             
 A sensible quantization procedure should respect the classical symmetry group of each orbit, the Virasoro group. This is the case of the geometric quantization scheme  \cite{Woodhouse} which  gives a unitary irreducible representation of the group of classical symmetries. 
Although coadjoint orbits of Virasoro group were extensively studied by the mathematics community (for instance see \cite{Roger} and references therein), an explicit realization of the geometric quantization procedure is missing. What is known about the quantization of some orbits comes from a representation  of the Virasoro algebra over a Hilbert space called Verma module. In fact, the partition function in the pure AdS$_3$ sector is usually derived from properties of the Verma module. The possible relation with the geometric quantization was first discussed in \cite{Witten88}.

Although the ultimate aim of our project is to address the open problems on quantization, this first calls for a detailed study of several classical aspects of the coadjoint orbits of Virasoro group. Because our interest is in the application to AdS$_3$ gravity, our exposition is oriented from the beginning to this theory and may somehow differ from the usual ones, cf. \cite{Witten88, Roger,Balog}. The description of at least a subset of the phase space of AdS$_3$ gravity by means of coadjoint orbits is not new and is explicit in \cite{Nakatsu, Navarro99}. Nevertheless, here we reconsider this approach and apply it exhaustively, with results that are valid globally in the space of solutions. Specifically, through the study of Virasoro orbits applied to AdS$_3$ gravity we find:

\begin{itemize}

\item The energy in the BTZ and AdS$_3$ orbits is bounded from below, being these geometries the ones with lowest energy in their respective orbits. This statement holds globally in the orbits.

\item An exhaustive classification of boundary gravitons apart from those of AdS$_3$ and BTZ black holes: \textit{exotic} boundary gravitons. We illustrate with examples.

\end{itemize}

In addition, we address the issue of extending away from infinity the geometries defined close to the boundary by showing that: 

\begin{itemize}

\item One of the exotic orbits has energy bounded from below and is included in the family of metrics with Killing horizons described in \cite{Escoceses}, which were shown to admit an extension beyond the horizon. 

\end{itemize}

The organization of this paper is the following: In Section 2, we briefly review the appearance of Diff$(S^1)$ and Virasoro symmetries in the context of three-dimensional gravity with a negative cosmological constant  with Brown-Henneaux boundary conditions. This presentation is tailor-made for what will come next, paying special attention to the status of the central element in the algebra of charges. Then, Section 3 is devoted to introduce the classification of coadjoint orbits of the Virasoro group in a concise and self-contained manner. In Section 4, we take advantage of the previous sections in order to revisit the description of the space of solutions of pure AdS$_3$ gravity. For example, we extend the description of the classical phase space of \cite{Castro} to any element of the AdS$_3$ orbit and also to other orbits of interest. In addition, we provide a simple proof of the boundedness from below of the energy for BTZ black holes, and by using previous results on Hills equation \cite{Balog}, also for AdS$_3$ and an exotic coadjoint orbit. This last orbit is shown to be close to the one of AdS$_3$ and also to be present in the recent work of \cite{Escoceses}. Finally, in Section 5, we discuss previous attempts to quantize the orbits and some ideas to explore in the future.

\section{Asymptotic solutions of gravity in $2+1$ dimensions}

 Let us consider the space of classical solutions of AdS$_3$ gravity. We want to emphasize that by AdS$_3$ gravity is meant a subset of Lorentzian metrics fulfilling Einstein equation with negative cosmological constant. This subset is comprised by those metrics that approach AdS$_3$ asymptotically. In what follows we use the common definition of the asymptotic condition given in \cite{BrownHenneaux}. The important remark that we want to make is that the introduction of this boundary condition breaks the full diffeomporphism invariance of general relativity. The boundary condition was motivated in \cite{BrownHenneaux} by the desire of having a well-defined variational principle. Also this allows to define finite conserved charges as surface integrals at the boundary.

\subsection{AdS$_3$ boundary conditions}

There is a rigorous definition of the asymptotic condition which can be given in a coordinate free language (see for instance \cite{Skenderis}). We will not use this language here. Following \cite{BrownHenneaux} we consider the family of spacetimes such that they are asymptotically AdS$_3$ in the following sense,
\begin{subequations}\label{amplitudes}
 \begin{align}
g_{+-}&=-\frac{r^2}{2} +\ell^2\gamma_{+-}(x^+,x^-)+\mathcal{O}(1/r),\\
g_{\pm\pm} &=\ell^2 \gamma_{\pm\pm}  (x^+,x^-) +\mathcal{O}(1/r),\\
g_{\pm\,r}&= \ell^4\frac{\gamma_{\pm\,r}(x^+,x^-)}{r^3}+ \mathcal{O}(1/r^4),\\
g_{rr}&= \frac{\ell^2}{r^2}+\ell^4\frac{\gamma_{rr}(x^+,x^-)}{r^4}+ \mathcal{O}(1/r^5).
\end{align}
\end{subequations}
We are using lightcone coordinates $x^\pm=t/\ell\pm \phi$ and $r$ is the radial coordinate. The boundary is at $r \rightarrow \infty$. The infinitesimal diffeomorphisms which leave invariant these asymptotic boundary conditions are given by vector fields $\zeta$ with components,
\begin{subequations}\label{asymptdiffs}
\begin{align}
\zeta^+&=\xi^++\frac{\ell^2}{2r^2}\partial_-^2 \xi^-+\mathcal{O}(1/r^4),\\
\zeta^-&=\xi^-+\frac{\ell^2}{2r^2}\partial_+^2 \xi^++\mathcal{O}(1/r^4),\\
\zeta^r&=-\frac{r}{2}(\partial_+ \xi^+ +\partial_- \xi^-) +\mathcal{O}(1/r),
\end{align}
\end{subequations}
where the functions $\xi^+$ and $\xi^-$ are chiral functions, namely they depend only on $x^+$ and $x^-$ respectively. 

\subsection{Proper and improper diffeomorphisms}

The vector field  of (\ref{asymptdiffs}) has two contributions. The first one is given by the terms containing the functions $\xi^+$ and $\xi^-$, which are explicitly displayed. The second one, is given by the higher order terms in powers of $1/r$. Following  \cite{BrownHenneaux}, when the first contribution vanishes the (infinitesimal) diffeomorphism is called a \textit{proper diffeomorphism}. Otherwise it is called an \textit{improper diffeomorphism}. This terminology should not be confused with the well-definiteness of the diffeomorphisms. The relevant distinction between the two contributions is in their functional dependence: while the proper diffeomorphism has an arbitrary dependence in the boundary coordinates, the improper one has a restricted dependence in the leading term since it is a sum of functions depending on one coordinate $x^+$ or $x^-$. Due to the arbitrary functional dependence, proper diffeomorphisms are treated as local gauge transformations. This implies that the metrics will be treated as physically equivalent only if they are related by a proper diffeomorphisms. On the other hand, improper diffeomorphisms are treated as global symmetries relating physically distinct metrics. This is the reason why the AdS$_3$ sector of \cite{MaloneyWitten} is not just the AdS$_3$ geometry, but contains metrics related to it by improper diffeomorphisms, i.e. its boundary gravitons.

\subsection{Exact asymptotic solution}

It can be shown that by making proper diffeomorphisms ($\xi^\pm=0$), any metric with the asymptotic conditions (\ref{amplitudes}) and  satisfying the equations of motion can be brought to the form \cite{Navarro98},
\begin{equation}\label{metricasympt}
ds^2=\frac{\ell^2}{r^2} dr^2-r^2dx^+dx^-+\ell^2\gamma_{++} (dx^+)^2+\ell^2\gamma_{--} (dx^-)^2+\ldots,
\end{equation}
where $\gamma_{\pm\pm}$ are now chiral functions and $\dots$ means sub-leading terms. The functions $\gamma_{\pm\pm}$  are invariant under proper diffeomorphisms and different sub-leading terms are gauge equivalent. In fact, there exists an exact solution parameterized by these chiral functions $\gamma_{\pm\pm}$ \cite{Banados},

\begin{equation}
ds^2=\frac{\ell^2}{r^2} dr^2-(r dx^- -\frac{\ell^2\gamma_{++}}{r}dx^+)(r dx^+ -\frac{\ell^ 2\gamma_{--}}{r}dx^-),\label{metricaexacta}
\end{equation}
where this radial coordinate is not the same as in (\ref{metricasympt}).

The important point to state now is that the infinitesimal diffeomorphisms generated by (\ref{asymptdiffs}) act on the functions $\gamma_{\pm\pm}$ as:
\begin{equation}
\delta_\xi^\pm \gamma_{\pm\pm} = 2\partial_\pm \xi^\pm \gamma_{\pm\pm} + \xi^\pm \partial_\pm \gamma_{\pm\pm}-\frac{1}{2}\partial_\pm^3 \xi^\pm\label{anomaloustrans}
\end{equation}

This is the familiar behaviour of the energy-momentum tensor $\Theta$ of a two-dimensional CFT under an infinitesimal conformal transformation. This is not an accident and fits into the AdS/CFT correspondence \cite{AdSCFT}. In fact,  the holographic stress tensor of AdS$_3$ \cite{BalasubramanianKraus} has components $\Theta_{\pm\pm}=\frac{\ell}{8\pi{G}}\gamma_{\pm\pm}$ and $\Theta_{-+}=\Theta_{+-}=0$,  in this coordinate system. Then, the anomalous term for $\Theta$ becomes $-\frac{1}{24\pi}\frac{3\ell}{2G}\partial_\pm^3 \xi^\pm$, with $\frac{3\ell}{2G}$ being the central charge of the expected dual CFT.  

\subsection*{Witt algebra: the algebra of asymptotic diffeomorphisms}

From now on we consider the equivalence class defined as the set of asymptotic vector fields (\ref{asymptdiffs}) modulo those generating proper diffeomorphisms. The elements of the equivalence class generate improper diffeomorphisms uniquely determined by the functions $\xi^{\pm}$. The Lie bracket of two such vector fields $\zeta_1$ and $\zeta_2$ is a new vector field $\zeta_{12}$ of parameter $\xi_{12}^{\pm}$ given by
$\xi^{\pm}_1({\xi^{\pm}_2})'-\xi^{\pm}_2({\xi^{\pm}_1})'$. 
So, the algebra of the asymptotic vector fields $\zeta$ determines two copies of a Lie algebra of the parameters $\xi^{+}$ and $\xi^{-}$, where the Lie product among them is defined by

\begin{eqnarray}
[\xi_1^{\pm},\xi_2^{\pm}]&:=&\xi_1^{\pm}({\xi_2^{\pm}})'-\xi_2^{\pm}({\xi_1^{\pm}})'\\
{[}\xi_1^{\pm},\xi_2^{\mp}]&:=&0
\end{eqnarray}

We will concentrate in only one copy. This algebra is  isomorphic to the Lie algebra of the group of diffeomorphisms of the circle called Vect$(S^1)$ (which will be explained in more detail in Section 3).  The interpretation of $\xi$ as a generator of a diffeomorphism of the circle comes from viewing $\xi$ as the component of the vector in the circle $ \xi\partial_\theta$, and now the previous bracket is well understood as the Lie bracket between vector fields on the circle. We will sometimes use $\xi$ meaning the vector $\xi \partial_\theta$.  Using the Fourier components $l_n(\theta)=ie^{in\theta}\partial_{\theta}$ of the vectors $\xi \partial_\theta$, the algebra product takes the usual form of the Witt algebra bracket:

\begin{equation}
[l_n,l_m]=(n-m)l_{m+n}
\end{equation}
We display this relations between complex combinations of the vectors fields in order to make contact with the literature. However, the relevant algebra for us will be the real algebra Vect$(S^1)$.

It is important to remark that the algebra of the inhomogeneous transformations of $\gamma$ is still Vect$(S^1)$, since this is a realization of the asymptotic diffeomporphisms algebra; the presence of the anomalous term in (\ref{anomaloustrans}) does not change this fact. More explicitly,

\begin{equation}
(\delta_{\xi_1}\delta_{\xi_2}-\delta_{\xi_2}\delta_{\xi_1})\gamma=\delta_{[\xi_1,\xi_2]}\gamma
\end{equation}

\subsection*{Virasoro algebra: the algebra of conserved charges}

Now, we will briefly review the well-known result of Brown-Henneaux \cite{BrownHenneaux}, that the algebra of the constraints of AdS$_3$ gravity is a central extension of the algebra of asymptotic symmetries Vect$(S^1)$. More precisely, in their analysis the Lie algebra product is the Poisson bracket between smeared constraints with the addition of a surface term. This surface term is what is called the charge. However, here we want to present this algebra from a different perspective, using only boundary data.

The conserved charges $Q_{\xi}[\gamma]$ are surface integrals determined by the parameter $\xi$ of the asymptotic diffeomorphism and the functions $\gamma$ which determine the boundary metric:
\begin{equation}\label{charges}
Q_{\xi^+,\xi^-}[\gamma_{++},\gamma_{--}]=\frac{\ell}{8\pi G}\int_0^{2\pi}dx^+ \xi^+ \gamma_{++}+\frac{\ell}{8\pi G}\int_0^{2\pi}dx^- \xi^- \gamma_{--}
\end{equation}
The separation between the $+$ and $-$ copies is manifest and so from now on we will refer to the charge of only one copy $Q_{\xi}[\gamma]$ without specifying which copy. The total charge is just the sum of both copies. With this expression for the charge, the BTZ of zero mass and angular momentum has null charge for any $\xi$. For the case of AdS$_3$ the mass is $-1/8G$ and all other charges vanish.

Brown and Henneaux, through their analysis, arrive to the Virasoro algebra. Their Lie bracket can be viewed, from a boundary point of view, as the following product between the linear functionals $Q_{\xi}[.]$,

\begin{equation}
\{Q_{\xi_1},Q_{\xi_2}\}[\gamma]:={-}Q_{\xi_2}[ \delta_{\xi_1}{\gamma}]\label{defchargealgebra}
\end{equation}

This bracket gives the variation of the charge $Q_{\xi_2}$ when the function $\gamma$ is perturbed by the infinitesimal diffeomorphism given by $-\xi_1$, according to (\ref{anomaloustrans}).
With this bracket, the following relation holds:

\begin{equation}
\{Q_{\xi_1},Q_{\xi_2}\}[\gamma]=Q_{[\xi_1,\xi_2]}[\gamma] + \frac{1}{48\pi} \,\frac{3\ell}{2G}\,  \int_{0}^{2\pi}(\xi_1'\xi_2''-\xi_2'\xi_1'')\label{chargealgebra}
\end{equation}
In particular, using the parameters $l_n=i e^{in\theta} $, the algebra of the charges is:

\begin{equation}
\{Q_{l_n},Q_{l_m}\}[\gamma]=(n-m)Q_{l_{n+m}}[\gamma]+i\frac{3\ell}{2G}\frac{n^3}{12}\delta_{n,-m}\label{chargealgebra}
\end{equation}

This product between charges, defined by (\ref{defchargealgebra}),  is actually not closed because of the presence of the  last constant term. In order to close the algebra product, we need to add to the family of charges $Q_{\xi}$ a constant  functional $K$. For convenience, and at this point this is completely arbitrary, we chose $K[\gamma]=3\ell/2G$, which is commonly known as the Brown-Henneaux central charge. When we do this, we recover the Virasoro algebra, which in a particular complex basis reads,
\begin{eqnarray}
[L_m,L_n]&=&(n-m)L_{n+m} + \frac{n^3}{12}Z \delta_{n,-m}\\
{[}Z,L_m]&=&0
\end{eqnarray}
This can be seen using the isomorphism:

\begin{eqnarray}
{L_m}\leftrightarrow Q_{l_n}[.]\\
{Z}\leftrightarrow iK[.]
\end{eqnarray}
and
\begin{equation}
\{K,Q_\xi\}:=0
\end{equation}

So far, we have presented two different kinds of functionals. The ones that come from (\ref{charges}) and the constant functional $K$. The former are labelled by the Vect$(S^1)$  elements $\xi$ while $K$ has no label. However, it will be convenient for later use to put all these functionals on equal footing.  To achieve this, we can consider the vector space of linear combinations of $L_m$'s and $Z$, in such a way that any charge can be written as $Q_{u}$, where $u$ belongs to this vector space and $Q$ is linear in $u$. The original charges are given by $Q_{L_m}:=Q_{l_n}$ while the constant functional is $Q_Z:=iK$. With this more democratic notation, the charge algebra reads,
\begin{eqnarray}
\{Q_u,Q_v\}=Q_{[u,v]}\label{BHalgebracomovir},
\end{eqnarray}
where the Lie bracket $[u,v]$ is the Virasoro Lie product.

\section{A primer on coadjoint orbits of Virasoro group}

In this section we review the basic features of the coadjoint representation of the Virasoro algebra which are important for the rest of the paper. We also clarify the notation and conventions we use.
\subsection{The coadjoint representation of Virasoro algebra}

Here we review the definition of the Virasoro algebra. Following \cite{Witten88,Schottenloher}, we first consider the group of orientation-preserving diffeomorphisms of the circle: ${\text{Diff}}_+(S^1)$ (from now on we omit the $+$ subscript). This is an infinite dimensional group and its real Lie algebra is $\text{Vect}(\text{S}^1)$, namely that of real vector fields $\vec{f}$ in the circle, defined by $\vec{f}\equiv{f} \partial_{\theta}$,  being the Lie product defined by the Lie bracket between vectors fields, i.e. $[\vec{f},\vec{g}]=(f g'-gf')\partial_{\theta}$. The central extension of Vect$(S^1)$ can be described by a pair $(\vec{f},a)$, being $\vec{f}$ a vector field and $a$ a real number. The Lie product is defined by,

\begin{equation}
\left[(\vec{f},a),(\vec{g},a')\right]=\left( [\vec{f},\vec{g}]\quad,\quad\frac{1}{48\pi} \int_{S^1}(f'g''-g'f'')\right).
\end{equation}

In order to make contact with the standard notation in physics, let us define ``the Virasoro modes'' by:
\begin{eqnarray}
L_n&:=&(l_n,0),\qquad\qquad l_n(\theta):=i e^{in\theta}\partial_{\theta}, \quad n\in \mathbb{Z},\nonumber\\
Z&:=&(0,i)\label{virasorobasis}
\end{eqnarray}
and the Lie product takes the usual form\footnote{It is more common to find the following definition:
$\tilde L_n=(l_n,0) \quad n\neq0$, $\tilde{L}_0=L_0 + \frac{1}{24}Z$. With this definition, the coefficient of $Z$ in the Lie bracket changes to $\frac{1}{12} n(n^2-1)\delta_{m+n,0}$.} ,
\begin{eqnarray}
[L_n,L_m]&=&(n-m) L_{m+n} +\frac{Z}{12} n^3\delta_{m+n,0}\\
{[}Z,L_n]&=&0
\end{eqnarray}
We are using a convention where there is no complex unit $i$ in the expression of the algebra product since it seems more simple in this way. However, note that the basis (\ref{virasorobasis}) is made of complex vectors, but nevertheless we will end up using real vector fields in the upcoming sections. Also, it is worth reminding the reader that by definition the Virasoro algebra, $vir$ from now on,  admits only finite combinations of the basis elements  in (\ref{virasorobasis}), while the central extension of Vect$(S^1)$ does not even refer to a  particular basis.

In the previous section we mentioned the Witt algebra. This is spanned by vector fields of the form
\begin{equation}
l_n(\theta)=i e^{in\theta}\partial_{\theta}, \quad n\in \mathbb{Z},
\end{equation}
and its Lie product is therefore given by
\begin{equation}
[l_n,l_m]=(n-m)l_{n+m}.
\end{equation}
Again, only finite linear combinations of these basis elements are admitted. Thus, $vir$ is the central extension of the Witt algebra. We will in general work with the algebra Vect$(S^1)$ and its central extension, but in order to avoid being pedantic we will sometimes make the abuse of calling them Witt algebra and Virasoro algebra respectively.  

\subsection{Adjoint and coadjoint representation of a Lie algebra}

Among all the representations of a Lie algebra $\mathfrak{g}$, the so-called {\it adjoint representation} is a very natural one since it assigns to each Lie algebra element $u$ a linear transformation $ad_u$ on the Lie algebra itself, as follows:

\begin{equation}
ad_u(v)\equiv{[}u,v]\,\qquad \forall v \in \mathfrak{g}.
\end{equation}
It is clear from this expression that the vector space where the Lie algebra elements act is indeed itself. This is a representation thanks to Jacobi identity.

There is another natural representation defined by the action of a Lie algebra element on its dual space $\mathfrak{g}^*$. Let us denote by $e^*$ a generic element of the dual space and by $<,>$ the pairing between $\mathfrak{g}^*$ and $\mathfrak{g}$. The coadjoint action $ad^{*}_v$ on an element $e^*$ is defined by the requirement that the pairing remains invariant under the action of the algebra:

\begin{equation}
<ad^*_v(e^*),u> + <e^*,ad_v u>=0\quad \Rightarrow \quad<ad^*_v(e^*),u>=-<e^*,[v,u]>.
\end{equation}

\subsection{Coadjoint representation of Virasoro algebra}

Let us start by considering the coadjoint representation of Vect$(S^1)$. This is the space of $(0,2)$-tensor fields $\tilde{b}$ in the circle given by $\tilde{b}=bd{\theta}^2$ (here $d\theta^2=d\theta \otimes d\theta$ and $b$ is a function on $S^1$) , whose pairing with the vector fields is defined as:

\begin{equation}\label{wittpairing}
< \tilde{b},\vec{f}>=\int_{S^1}bf
\end{equation}

In the same manner a pair $(\vec{f},a)$ can be used to refer to an element of $vir$, a pair $(\tilde{b},t)$, with $t$  a real number, serves to characterize an element in the dual space of the Virasoro algebra $vir^*$, where the pairing is now defined by:

\begin{equation}
< (\tilde{b},t),(\vec{g},a) >=\int_0^{2\pi}d\theta g(\theta) b(\theta)+a t
\end{equation}
The coadjoint action is given by
\begin{equation}\label{coadjointaction}
\text{ad}^*_{(\vec{f},a)}(\tilde{b},t)=\left(\left(2f'b+fb'-\frac{t}{24\pi}f'''\right)d\theta^2,0  \right),
\end{equation}
which leaves the pairing of adjoint and coadjoint vectors of $vir$ invariant. It is important to remark that the coadjoint action of $vir$, viewed as an infinitesimal change in $vir^*$, does not change the central element: $\delta t=0$. Equally important is the fact that the coadjoint representation of the central element $Z$ is the null endomorphism, $ad^*_Z=0$. In other words, this is a representation of zero central charge. Because of this, the coadjoint action (\ref{coadjointaction}) of $vir$ can be viewed as a realization of Witt algebra, neglecting the second argument of $(\vec f,a)$ above, namely keeping only $\vec f$.

\subsection{Coadjoint orbits} 

In order to describe what is known as coadjoint orbits of a group, we need first to introduce the adjoint and coadjoint actions of the group. Let us give a brief general summary and then look at the case of the group Diff$(S^1)$ and its central extension (for a more detailed description see for example \cite{Roger}). The adjoint action of a Lie group $G$ is a group representation $Ad: G \rightarrow \text{Aut}(\mathfrak{g})$ over its Lie algebra.  The coadjoint action of the group is a representation that acts on the dual space of the Lie algebra, namely $Ad^*: G \rightarrow \text{Aut}(\mathfrak{g}^*)$. If we denote again the pairing of an adjoint vector $v$ and a coadjoint vector $e^*$ by $<e^*,v>$, then the coadjoint action of the group must satisfy $<Ad_g^* e^*,Ad_gv>=<e^*,v>$, and so the coadjoint group action $Ad^*_g $ is the transpose of the adjoint action with the inverse group element $^t(Ad_{g^{-1}})$.

For the case of $G=\text{Diff}(S^1)$, the adjoint group action is the push-forward on vector fields on $S^1$ by  elements of Diff$(S^1)$, namely, if $g \in \text{Diff}(S^1)$, then $Ad_g(v)=g_*(v)$ for any vector field $v$ in $S^1$. Explicitly, if $s:\theta \mapsto s(\theta)$ is a diffeomorphism of the circle, then for any $f\partial_\theta$ we have that the vector push-forward has a new component $f_s$ given by 
\begin{equation}
 f_s(\theta)=\frac{(f \circ s)(\theta)}{s'(\theta)}.
 \end{equation} 

The coadjoint group action, as explained above, is the transpose of the pushforward by the element $g^{-1}$. So, if $\theta\mapsto s(\theta)$ is the diffeomorphism, then $b$ is mapped to $b_s$ as,
\begin{equation}
 b_s=(b \circ s)s'^2   .
 \end{equation}
This makes the pairing $<\tilde b,\vec f>$ defined in (\ref{wittpairing}) invariant. The coadjoint action of $\text{Diff}(S^1)$ can be used to construct the coadjoint group action for the central extension of Diff$(S^1)$  \cite{Roger}. This central extension is the Virasoro group which we will call $Vir$. The construction takes into account that the central element is trivial under $Ad^*$ and so $Ad^*:\text{Diff}(S^1)\times vir^ *\rightarrow vir^ *$ in the following way,
\begin{equation}
 (b_s d\theta^2,t_s)=\left((b\circ s) s'^2d\theta^2-\frac{t}{24\pi}S(s),t\right),
\end{equation}
where
\begin{equation}
S(s)=\frac{s'''}{s'}-\frac{3}{2}\frac{s''^2}{s'^2}
\end{equation}
is the Schwarzian derivative. The infinitesimal coadjoint action is precisely (\ref{coadjointaction}).

Now we are in a position to define a coadjoint orbit of a dual vector $e^*$,
\begin{equation}
W_{e^*}=\{r^* \in \mathfrak{g}^*/ \quad r^*=Ad^*_ge^*, \quad g\in G\}.
\end{equation}
In particular, the coadjoint orbit of an element $(\tilde{b},t) \in vir^*$ is defined as
\begin{equation}
W_{(\tilde{b},t)}=\{(\tilde{b}',t') \in vir^* / \,\, (\tilde{b}',t)=\text{Ad}^*_{g}(\tilde{b},t),\quad g\in Vir\},
\end{equation}
so it is the image of the coadjoint action of the Virasoro group on the coadjoint vector $(\tilde{b},t)$.

Coadjoint orbits have a manifold structure, being isomorphic to the homogeneous space $G/H$, where $G$ is the Lie group and $H$ the stabilizer of $(\tilde{b},t)$. Namely, in our case $G=Vir$ and
\begin{equation}
H=\{g \in Vir /\quad Ad^*_g (\tilde{b},t)=(\tilde{b},t) \}
\end{equation}
One should be more precise and put a subscript in $H$ that indicates the coadjoint vector that is left invariant by the elements of $H$: $H_{(\tilde{b},t)}$. Although we may avoid this notation, it is important to keep in mind that there are several different subgroups $H$ of $Vir$. We will discuss these possibilities next, but first a cautionary note: the isotropy groups of Virasoro will be of the form $\mathbb{R} \times \tilde{H}$, where the first factor refers to the center of the group (whose algebra is generated by the central element $Z$), which is trivial under the coadjoint representation. On the other hand, the number $t$ in any coadjoint vector $(\tilde{b},t)$ is always left invariant by the coadjoint action. This means that the orbits $Vir/(\mathbb{R} \times \tilde{H})$ and Diff$(S^1)/\tilde{H}$ are the same. For this reason it suffices to study the orbits Diff$(S^1)/H$, with $H$ a subgroup of Diff$(S^1)$.

\subsubsection*{Classification of Virasoro orbits}

For reasons that will become clear in Section 4, we will be firstly interested in the orbit of a ``constant'' coadjoint vector $(b_0d\theta^2,t)$, with $b_0$ a real number. Generically, the group $H$ that leaves this coadjoint vector invariant is generated by $\partial_\theta$, and can be identified with $S^1$. Namely, the coadjoint orbit of $(b_0d\theta^2,t)$ is
\begin{equation}
     W_{(b_0d\theta^2,t)}=\frac{Vir}{(\mathbb{R}\times S^1)}=\frac{\text{Diff}(S^1)}{S^1}.
\end{equation}
We say generically because there are special constant points: when $b_0=-n^2t/48\pi,\quad n\in \mathbb{N}$, the stabilizer is generated\footnote{In order to use only real vector fields we should say $\{\partial_\theta,\cos(n\theta)\partial_\theta,\sin(n \theta)\partial_\theta\}$, but using the Virasoro modes allows to keep contact with the usual notation in the literature.} by $\{l_0,l_n,l_{-n}\}$ and the group is $H=\text{PSL}_2^{(n)}$ (an $n-$covering of PSL$_2$). The case $n=1$ will be of importance to us later. The orbits with $(b_0 d\theta^2,t)$ points, i.e. with constant representatives, are the ones we have mentioned so far. We have a simple argument why these orbits have a unique constant representative, for the case when this is positive. This is explained in Section 4. In fact  this result is valid for all values of the constant representative, not only for the positive ones \cite{Balog}. So, there is at most one constant representative in an orbit, and the orbits with no constant representatives are introduced next.

It can be proved (see \cite{Witten88} for example) that the algebra of $H$ is always one- or three-dimensional. The orbits that we will discuss now have a one-parameter isotropy group: $T_{n,\Delta}$ or $\tilde{T}_{n,\pm}$ \cite{Witten88,Balog}, so in both types $H$ is a group of dimension one. The first stabilizer group\footnote{In \cite{Balog} they call $L$ what we call $\gamma$, and $4\pi b$ what we call $\Delta$.}, $T_{n,\Delta}$, is generated by a vector field $f\partial_\theta$ with $2n$ simple zeros  ($n\neq 0$) and with $|f'|=1$ at each zero \cite{Witten88}. This orbit possesses the following orbit invariant,
\begin{equation}\label{Delta}
\Delta:=\lim_{\epsilon\rightarrow 0}\int_{S^1-V_{\epsilon}}\frac{1}{f},
\end{equation}
where $V_{\epsilon}$ are open sets of volume $\epsilon$ around the zeros of $f$, i.e., $V_\epsilon$ is the union of the intervals $(x_k-\epsilon,x_k+\epsilon)$, with $x_k$ the zeros of $f$. It turns out that $\Delta$ together with $n$ are sufficient to characterize this kind of orbits. These orbits have no constant representative\footnote{If it had a constant representative, it should be stabilized by the $l_0$ vector, but this vector has no zeros. Since the number of zeros is left invariant, $l_0$ cannot be the push-forward of $f\partial_\theta$.} but tend to the orbits with isotropy group PSL$_2^{(n)}$ as $\Delta\rightarrow 0$.  In this sense the orbits Diff$(S^1)/T_{n,\Delta}$ are perturbations of the orbits Diff$(S^1)/$PSL$_2^{(n)}$ for fixed $n$.

The last orbits are the ones with isotropy groups $\tilde{T}_{n,\pm}$, which are generated by vectors $\tilde{f_{\pm}}\partial_\theta$ with $n$ double zeros and with null third derivative at each zero \cite{Witten88}. The $\pm$ sign indicates two inequivalent orbits with isotropy group generated by these vectors, i.e. this sign is an orbit invariant but its calculation is not relevant here. In \cite{Witten88} these orbits are said to converge to the orbits Diff$(S^1)/$PSL$_2^{(n)}$ too. In \cite{Balog} an explicit one-parameter curve inside an orbit Diff$(S^1)/\tilde{T}_{n,\pm}$  is constructed and it is shown there that in some limit the points in the curve approach the constant representative of Diff$(S^1)/$PSL$_2^{(n)}$, for fixed $n$ and $\pm$ sign.

Let us summarize the information we have on the different coadjoint orbits of Virasoro group in the following table
$$$$
\begin{center}
\begin{tabular}{|c|c|c|}
  \hline
  Orbit & Algebra stabilizers & Representative  \\ \hline

  Diff$(S^1)/S^1$ & $\{l_0\}$ & $b_0\neq -tn^2/48\pi $ and constant \\ \hline

  Diff$(S^1)/$PSL$^{(n)}_2$ & $\{l_0,l_n,l_{-n}\}$ & $b_0=-tn^2/48\pi $   \\ \hline
    
  Diff$(S^1)/T_{n,\Delta}$ & \parbox[t]{5cm}{$\{f\partial_\theta\}$, $f$ has $2n>0$ simple zeros. $n$ and $\Delta$ are invariants.}  & See (\ref{bforTnDelta}) in Section 4\\ \hline

 Diff$(S^1)/\tilde{T}_{n,\pm}$ & \parbox[t]{5cm}{$\{f\partial_\theta\}$, $f$ has $n$ double zeros\\$n$ and $\pm$ sign are invariants.} & See  (\ref{bforTnmasmenos}) in Section 4\\
  \hline
\end{tabular}\\
\vspace{5mm}{\small Table 1. Classification of Virasoro coadjoint orbits.} 
\end{center}

\subsection{Symplectic form in the coadjoint orbits}

Probably one the most important features of coadjoint orbits is the fact that always have a $G-$invariant symplectic structure. Let us first consider a coadjoint vector $e^*$ and a two form $\omega_{e^*}$ over the orbit $W_{e^*}$. We can identify an element of the tangent space of the orbit at $e^*$, $T_{e^*}W_{e^*}$, with an element of the algebra $\mathfrak{g}$ as follows: any element of the tangent space at $e^*$ will be of the form $ad^*_u e^*$, with $u \in \mathfrak{g}$ being defined up to elements of the Lie algebra of the isotropy group $H_{e^*}$ (i.e., elements $v \in \mathfrak{g}$ such that $ad^*_v e^*=0$). So, for any element $\bar{u} \in T_{e^*}W_{e^*}$ we have an element $u \in \mathfrak{g}$ up to adjoint vectors that leave  $e^*$ invariant.

The symplectic form is defined by
\begin{equation}\label{symplecticform}
 \omega_{e^*}(ad^*_u e^*, ad^*_v e^* )=< e^*, [u,v] >.
\end{equation}
It can be shown that this is well defined: for any two $u$ and $u'$ in $\mathfrak{g}$ such that $ad^*_u e^*=ad^*_{u'} e^*$ and $u \neq u'$, the previous expression does not change. This symplectic form defined at $e^*$ can be consistently extended to the whole orbit by the action of $G$, therefore being $G$-invariant. Moreover, it is a non-degenerate two form. Thus, $W_{e^*}$ is a symplectic manifold with $\omega$ being the symplectic form.

\subsection{Poisson bracket in each orbit}

Now, let us discuss the Poisson structure induced by the symplectic form we just introduced. For an arbitrary coadjoint vector $e^*$ we have the corresponding coadjoint orbit $W_{e^*}$, and for any Virasoro element $u \in vir$  a function $\Phi_u:W_{e^*} \rightarrow \mathbb{R}$ can be defined as
\begin{equation}\label{Phifunctions}
\Phi_u(r^*)=< r^*,u >,
\end{equation}
for any $r^*\in{W_{e^*}}$. If needed, we will extend their image to the complex plane by linearity.
These functions are the ones that will benefit from the symplectic structure on $W_{e^*}$. By means of the inverse of the symplectic form, a Poisson bracket $\{,\}_{\omega^{-1}}$ can be defined. Moreover, the functions $\Phi_u$ will realize the algebra, in our case the Virasoro algebra, through the Poisson bracket,
\begin{equation}\label{poissonvirasoro}
\{\Phi_u,\Phi_v\}_{\omega^{-1}}=\Phi_{[u,v]},\qquad u,v\in vir.
\end{equation}
Later we will comment in detail the relation (an equality actually!) between these functions and the Brown-Henneaux charges. 

Consider any element  $(\tilde{b},t)$ in the dual algebra $vir^*$, then the central element $Z=(0,i)$ assigns to it the value $t$ through the function\footnote{Recall that we extend by linearity the image of $\Phi_u$ to the complex plane, when $u$ is a complex vector.} $\Phi_Z$:
\begin{equation}
\Phi_{Z}(\tilde{b},t)=< (\tilde{b},t), Z > = it
\end{equation}
Moreover, since the pairing $<,>$ is invariant under the coadjoint action of the algebra and so is $Z$, this assignment holds for the entire orbit $W_{(\tilde{b},t)}$. Thus, we can talk about the ``central charge'' of an orbit meaning (the imaginary part of) $\Phi_{Z}$ evaluated at any point of the coadjoint orbit. Because of this, from now on we will refer to $t$ as the central charge of an orbit. This is precisely the same map we had in the AdS$_3$ gravity analysis of the previous section, where we mapped $Z$ to the constant functional $iK$, and $K$ takes the value $t$ which is commonly referred to as the central charge.

A last comment is in order: the realization of the Virasoro algebra by means of the Poisson bracket (\ref{poissonvirasoro}) sends $Z$ to $\Phi_Z=it$ for a specific coadjoint orbit. On the other hand, by means of the symplectic structure, we can define Hamiltonian vector fields for each $\Phi_u$ with $u \in vir$. The Hamiltonian vector field associated to $\Phi_Z$, let us call it $\delta_Z$, is precisely a null vector since as we just saw $\Phi_Z$ is a constant function. This means that $Z$ generates no infinitesimal transformation ($\delta_Z=0$) on the functions, i.e. no change on classical observables  over the coadjoint orbits. However, we will later see that, upon quantization, $Z$ will have a non-trivial action on the space of quantum states.

\subsection{Bounded energy}

It is an interesting question which of the coadjoint orbits of Virasoro have an energy bounded from below . By energy of a generic point $B:=(bd\theta^2,t)$ we actually mean the value
\begin{equation}\label{energydefinition}
E(B)=-i\Phi_{L_0}(B)=\int_{S^1}b,
\end{equation}
or in other words $2\pi$ times the zero mode of $b$. The explicit analysis of \cite{Balog} shows that the orbits with a global minimum of the energy can be divided in two categories:   the first one is defined by the orbits with a constant representative such that
\begin{equation}\label{b0bound}
b_0 \geq -t/48\pi.
\end{equation}
The minimum energy, $2\pi b_0$, is reached precisely at the constant representative. These orbits are all of the form Diff$(S^1)/S^1$ except for the equality in (\ref{b0bound}), where the orbit is Diff$(S^1)/$PSL$_2$.

The other category contains only the orbit with isotropy group $\tilde{T}_{1,+}$, which has no constant representative. The largest minimum energy is $-t/24$, but is never reached along the orbit. This is intimately related with the fact that this orbit is, in a sense, a perturbation of Diff$(S^1)/$PSL$_2$, whose constant representative does have such minimum energy.

\section{AdS$_3$ gravity as a collection of orbits}

In Section 2 we studied the phase space of asymptotically AdS$_3$ solutions which is parameterized by two periodic functions $\gamma_{++}(x^+)$ and $\gamma_{--}(x^-)$. It was shown there that the change in a function $\gamma$  generated by an improper asymptotic diffeomorphism of the form (\ref{asymptdiffs}) realizes the Witt algebra but also can be seen as the coadjoint action of Virasoro algebra, once a central element is added to parameterize the phase space. By this we mean that now the pair $(\Theta,t)$ labels (one of the copies of) a particular gravitational solution, where $\Theta=\frac{\ell}{8\pi G}\gamma$ and $t$ is certain real number. This number is fixed by the transformation behaviour in (\ref{anomaloustrans}) to $t=\frac{3\ell}{2G}$.

From what we saw in Section 3, it should be clear that the space of metrics of Section 2 is a collection of coadjoint orbits of the Virasoro group. The orbit associated to a given metric $g_0$ is generated by the application of an arbitrary improper diffeomorphism to $g_0$. Each of these orbits can be endowed with a symplectic structure which is Virasoro-invariant, so they are symplectic manifolds and the Virasoro action acts as symplectomorphisms. In addition, and of most importance, the functions on the orbit $\Phi_u$ defined in (\ref{Phifunctions}) are the conserved charges $Q_u$, taking values on the orbit and with their Poisson bracket being homomorphic to the Virasoro algebra. The number $t$ is the imaginary part of the value that the central element takes on this realization, i.e. $Z \mapsto it$Id. This number $3\ell/2G$ is the known Brown-Henneaux central charge, that we will denote as usual by $c$.

In this section we use what we learned from the study of the Virasoro orbits in order to characterize the asymptotic metrics of physical relevance. For this purpose we analyse the space of the functions $\gamma$ which describe the metrics close to the boundary. This allows to arrive to interesting results, and an important one is the existence of energy bounds on the space of AdS$_3$ solutions. But first, two remarks are in order. In the first place, the expression (\ref{metricaexacta}) is a solution of Einstein-Hilbert equations with negative cosmological constant for any pair of smooth periodic functions $(\gamma_{++}$ , $\gamma_{--})$ and for $r>r_0$ (with $r_0$ a non-negative number). However, it could happen that the metric is not acceptable from the physical point of view, due to the possible presence of closed timelike curves or naked singularities, once it is continued to $r<r_0$. In addition, there could be in principle more than one possible extension of the metric. These observations show that the space of pairs $(\gamma_{++},\gamma_{--})$ is not a priori identical to the space of metrics of physical interest\footnote{An interesting work on the physical phase space of AdS$_3$ gravity is \cite{KrasnovScarinci}}. Nevertheless, we will see at the end of the present section that in fact some orbits contain geometries that admit an acceptable extension passing through a degenerate Killing horizon.

In the second place, let us suppose that   $(\gamma_{++}$, $\gamma_{--})$ does define a physically acceptable metric and let us consider their orbits. Then, the  question is if the other pairs of functions $\gamma$ in the orbits define physically acceptable metrics. It is clear that close to the boundary this is the case since they are all related by an asymptotic diffeomorphism. A problem could appear if one of these asymptotic diffeomorphisms cannot be extended far from the boundary. However this seems hard to be the case. Then, if one representative of an orbit has an acceptable extended metric we may assume that the other elements in the orbit also posses an acceptable extension. In any case we do not need to address this issue in what follows.

\subsection{The orbits of AdS$_3$ and BTZ black holes}
 
First of all, we are interested in the orbits of AdS$_3$ and of BTZ black holes. There will be a coadjoint vector which can be identified with these geometries. One comment concerning normalization. We describe the geometries by the pair $(\Theta,c)$ (instead of any other re-scaled version $(k\Theta,kc)$) in order that the canonical charges $Q_{L_n}$ defined in (\ref{charges}) in the gravity context agree with the functions $\Phi_{L_n}$.

Let us start with AdS$_3$. From (\ref{metricaexacta}) we know that AdS$_3$ corresponds to $\gamma_{++}=\gamma_{--}=-\frac{1}{4}$ and so

\begin{equation}\label{Ads3orbits}
\text{AdS}_3 \leftrightarrow \left(-\frac{\ell}{32\pi G} d\theta^2,  \frac{3\ell}{2G}\right)\times \left(-\frac{\ell}{32\pi G} d\theta^2,\frac{3\ell}{2G}\right)
\end{equation}
We see then that both coadjoint vectors, which together are identified with AdS$_3$, belong to the $n=1$ orbit Diff$(S^1)/$PSL$^{}_{2}$. This orbit has a constant representative $b_0=-c/48\pi$ and this is precisely the one needed for AdS$_3$. As we saw, the Lie algebra of its isotropy group is generated by $\{l_{-1},l_0,l_1\}$, which agree with the known isometries of one of the copies of AdS$_3$. The only non-zero charge is the mass, which by definition is the charge associated to the vector $\partial_t=1/\ell\,\partial_++1/\ell\, \partial_-$:
\begin{equation}
M=\Phi_{(\partial_t,0)}[AdS_3]=-\frac{i}{\ell}\Phi_{L_0}[(-c/48\pi, t)]-\frac{i}{\ell}\Phi_{\bar{L}_0}[(-c/48\pi, t)]=-\frac{c}{12\ell}.
\end{equation}
This agrees with the value of the mass $-1/8G$ obtained from the definition of the charge (\ref{charges}) with $c=3\ell/2G$.

Now we turn to the BTZ black holes, which are defined by the pair 
\begin{equation}
\frac{\ell}{8\pi{G}}\gamma_{++}=b_0\geq{0},\qquad \frac{\ell}{8\pi{G}}\gamma_{--}=\bar{b}_0\geq{0},
\end{equation}
being $b_0$ and $\bar{b}_0$ constant values. Thus
\begin{equation}
\text{BTZ} \leftrightarrow (b_0d\theta^2, c)\times(\bar{b}_0 d\theta^2, c)
\end{equation}
The constants $b_0$ and $\bar{b}_0$ are related to the mass $Q_{\partial_t}$ and angular momentum $Q_{\partial_\phi}$ by:
\begin{equation}
M=2\pi\frac{b_0+\bar{b}_0}{\ell},\quad J=2\pi(b_0-\bar{b}_0)
\end{equation}

As in the case of AdS$_3$, both coadjoint vectors describing BTZ geometries are constant but in this case belong to the generic orbit Diff$(S^1)/S^1\times\text{Diff}(S^1)/S^1$. As we pointed out, for the BTZ these constant representatives are non-negative. We saw that their isotropy group has a Lie algebra generated by $l_0$ and $\bar{l}_0$ respectively, which agree with the known isometries of BTZ black holes.
From the results in Section 3 we can reach important conclusions about the orbits that host AdS$_3$ and BTZ black holes:

\subsubsection*{Non-diffeomorphic spacetimes} 

Each orbit of the type Diff$(S^1)/S^1$ and Diff$(S^1)/$PSL$^{(n)}_2$ contains only one constant representative. This implies that BTZ geometries with different values of $M$ and $J$ must belong to different orbits and then cannot be related by improper diffeomorphisms; this also implies that AdS$_3$ is not related with BTZ black holes by improper diffeomporphisms.

We want to recall that the definition of a metric in terms of $\gamma$'s and the improper diffeomorphisms are defined for large values of $r$. So, say we we have two $\gamma$'s defining an asymptotic metric, this does not imply that there will exist a globally defined metric having an asymptotic behaviour characterized by the given $\gamma$'s. It could also happen that the improper diffeomporphism relating two points of a given orbit could not be extended to any value of $r$. However, we are using the argument in the opposite direction and we conclude that different BTZ black holes and AdS$_3$ are not related by diffeomorphisms even in the asymptotic region.

\subsubsection*{Boundedness from below of the energy for BTZ and AdS$_3$ orbits}

We are going to show that the energy in the BTZ and AdS$_3$ orbits is bounded from below and these well-known geometries are precisely the only ones that saturate the bound \footnote{Soon after we submitted the first version of this paper, a work appeared regarding energy bounds which arrives to the same results as here \cite{BarnichOblak}.}. The conclusions reached in this subsection are valid for any metric of the physical phase space\footnote{In \cite{Nakatsu} the energy bound of AdS$_3$ and BTZ orbits was studied although locally around AdS$_3$ and BTZ geometries. Here we show that the bounds remain valid globally and also consider the other kinds of orbits.}, since only their asymptotic behaviour is what matters to compute their energy.

Recall the definition of energy (\ref{energydefinition}) as a function on a given orbit and that the mass of a given metric is just the sum of the energies in each copy of coadjoint orbits (modulo an $\ell$ factor). We found a simple argument that shows that, at least for the orbits of BTZ geometries, the value of the energy of $(bd\theta^2,c)$ is greater than or equal to the one of the $(b_0 d\theta^2,c)$, being $b_0$ the constant representative. In other words, this result implies that each metric in the orbit of a BTZ of mass $M$ will have a mass grater than $M$. This can be proved by similar arguments as in \cite{Balog}. We know that $(b_2d\theta^2,c)$ is in the same orbit as $(b_1 d\theta^2,c)$ if and only if there exists a diffeomorphism given by a function $F$ such that:

\begin{equation}\label{finitediff}
b_2 (\theta) = F'^2(\theta)b_1 (F(\theta)) -\frac{c}{24\pi}\left(\frac{F'''}{F'}-\frac{3}{2}\frac{F''^2}{F'^2}\right)
\end{equation}

Let us consider the case in which $b_1$ is a non-negative constant (so it is one copy of a BTZ geometry) and let us compute the zero mode of the r.h.s of (\ref{finitediff}). Since the factor in front of $b_1$ is positive and $b_1\geq{0}$, the zero mode of the first term will be non-negative. On the other hand, using that $\frac{F'''}{F'}=(\frac{F''}{F'})'+ \frac{F''^2}{F'^2}$, we can see that the integral of the second term will be grater or equal than zero (since $\int (\frac{F''}{F'})'=0$). We then conclude,
\begin{equation}
\int_{0}^{2\pi}b_2(\theta)d\theta \geq 0
\end{equation}
By noticing\footnote{To see this write $F=\theta + \sum_{n \in \mathbb{Z}}a_n e^{in\theta}$ and note that the zero mode of $F'$ is $1$, so the zero mode of $F'^2$ must be $1+2 \sum_{n>0} |a_n|^2$.} that $\int_{S^1} (F')^2 \geq{2\pi}$, we can strengthen the inequality to show that:

\begin{equation}\label{inequality}
\frac{1}{2\pi}\int_{0}^{2\pi}b_2(\theta)d\theta\geq{b_1}
\end{equation}
This means that not only the energy is bounded from below but also that its lowest value is that of the constant representative. In the case of  a BTZ orbit, the minimum energy is just the BTZ mass $M$. Even more, it can be shown that it is a minimum strictly, i.e. that all the boundary gravitons associated to that BTZ geometry have  greater energy. As a corollary, we find that there is only one non-negative constant representative in each orbit: suppose now that $b_2$ is constant as well, then we would have $b_2 \geq b_1$, but with the inverse diffeomorphism we would arrive to $b_1 \geq b_2$, implying that the only consistent possibility is $b_2=b_1$.

For the case of AdS$_3$, whose constant representative is negative, we have to use the results of Section 3. There, we saw that any orbit with constant representative $b_0\geq -c/48\pi$ has a global minimum energy and that this minimum is $2\pi b_0$, i.e. reached only when $b=b_0$ \cite{Balog}. This means that, recalling (\ref{Ads3orbits}), AdS$_3$ is the metric with strictly the lowest energy in the orbit it belongs to. This fact holds for any metric identified with constant representatives such that $b_0\geq -c/48\pi$.  

Let us turn to the other orbits with no constant representative or with a constant representative such that $b_0 < -c/48\pi$. We can say, based on Section 3, that their energy is not bounded from below, except for the orbit Diff$(S^1)/  \tilde{T}_{1,+}$: it has a lower bound which is never reached, given by $-c/24$. We will comment more on this in what follows.
 
\subsection{Exotic boundary gravitons}

Apart from the BTZ orbits (orbits of the type Diff$(S^1)/S^1\times$Diff$(S^1)/S^1$ with non-negative constant representative) and the AdS$_3$ orbit (Diff$(S^1)/$PSL$_2\times$Diff$(S^1)/$PSL$_2$), there are also products of Virasoro orbits of the following type:

\begin{enumerate}
\item Diff$(S^1)/S^1\times$Diff$(S^1)/S^1$, with a negative constant representative in one copy or in both 

\item Diff$(S^1)/S^1\times$Diff$(S^1)/$PSL$^{(n)}_2$, 

\item Diff$(S^1)/$PSL$^{(m)}_2\times$Diff$(S^1)/$PSL$^{(n)}_2$, with $n$ or $m$ greater than 1. 
\end{enumerate}
These cases describe BTZ-like geometries with $M<|J|$, and according to \cite{MiskovicZanelli} they  posses topological defects or naked singularities.

On the other hand, we also have the orbits with isotropy groups $T_{n,\Delta}$ and $\tilde{T}_{n,\pm}$. Let us concentrate in the first ones now. $T_{n,\Delta}$ : these orbits do not have a constant representative. Fortunately, in \cite{Balog}  an explicit representative $b_{n,\Delta}$ was found on these orbits
\begin{equation}\label{bforTnDelta}
\frac{12\pi}{t}b_{n,\Delta}=\frac{\Delta^2}{(4\pi)^2} + \frac{n^2+\frac{\Delta^2}{4\pi^2}}{2Y}-\frac{3n^2}{4Y^2}
\end{equation}
with 
\begin{equation}
Y(\theta):=\cos^2\left(\frac{n\theta}{2}\right) + \left(\sin\left(\frac{n\theta}{2}\right)+ \frac{\Delta}{2n\pi}\cos\left(\frac{n\theta}{2}\right) \right)^2
\end{equation}
As was stated in Section 3, the numbers $\Delta$ and $n$ fully characterize the orbit, so there is always one and only one element of the form (\ref{bforTnDelta}) in each orbit. 

Now we turn to the other orbits with isotropy group $\tilde{T}_{n,\pm}$: these have no constant representative either. Again, there is an explicit element of the orbit found in \cite{Balog}
\begin{equation}\label{bforTnmasmenos}
\frac{12\pi}{t}b_{n,\pm}= \frac{n^2}{2\tilde{Y}}-\frac{3n^2(1-\frac{\pm 1}{2\pi})}{4\tilde{Y}^2}
\end{equation}
with 
\begin{equation}
\tilde{Y}(\theta):=1-\frac{\pm 1}{2\pi}\sin^2\left(\frac{n\theta}{2}\right)
\end{equation}
The number $n$ together with the sign $\pm$ define these orbits and there is one and only one element of the form (\ref{bforTnmasmenos}) in each orbit of this kind. 

In any orbit with isotropy groups $T_{n,\Delta}$ and $\tilde{T}_{n,\pm}$ one can construct a curve in the orbit. We will refer now to the latter (see equation (4.34) in \cite{Balog} for an example in the other case). An example of a curve in Diff$(S^1)/\tilde T_{n,\pm}$ is
\begin{equation}\label{Tmonocurve}
 \frac{12\pi}{t}b_{n,\pm}(a)=\frac{n^2a^2}{2\tilde{Y_a}}-\frac{3n^2a^2(a^2-\frac{\pm 1}{2\pi})}{4\tilde{Y_a}^2}, \qquad a\in \left(\frac{1}{\sqrt{2\pi}},\infty\right)
\end{equation}
with 
\begin{equation}
\tilde{Y}_a(\theta):=a^2-\frac{\pm 1}{2\pi}\sin^2\left(\frac{n\theta}{2}\right).
\end{equation}
In the limit $a\rightarrow\infty$ the curve approaches $b=-n^2c/48\pi$. So it is clear that these orbits are ``close'' at some point to the constant representative of the Diff$(S^1)/$PSL$_2^{(n)}$ orbits. Let us consider the case $n=1$ and $+$ now, namely the orbit Diff$(S^1)/  \tilde{T}_{1,+}$. We saw in Section 3 that this particular orbit has the energy bounded from below, but never reaches the bound $-c/24$. Now it is more clear what is happening: the elements of the curve (\ref{Tmonocurve}) approach the constant representative $b_0=-c/48\pi$ and so they get closer and closer to its energy but never reach it. In the gravity picture, this means that there is a set of metrics which are continuously connected to AdS$_3$, with strictly greater energy, and are non-diffeomorphic to it even in the asymptotic region.

\subsection{Diffeomorphic invariants}

 Let us consider here the problem of characterizing the sets of metrics related by diffeomporphisms. For two choices of the pair $(\gamma_{++},\gamma_{--})$, the corresponding boundary metrics could be related by an improper diffeomorphism or not. In other words, the corresponding $b$'s will be in the same Virasoro orbit or not. A natural question that we want to address is: given two elements $(b_1d\theta^2,t)$ and $(b_2d\theta^2,t)$, how can we decide whether they belong to the same orbit or not?

A necessary and sufficient condition for $b_1$ and $b_2$ to belong to the same orbit is the existence of a diffeomorphism given by $F$ such that (\ref{finitediff}) is satisfied. However,  since the existence of $F$ requires from us to be able to show that a complicated differential equation admits a regular solution with no zeros, this is not a practical criterion.

It would be useful to have a diffeomorphism-invariant characterization of the $b$'s by numbers  defining the orbits in a unique way. The canonical charges are not useful, since these are just the Fourier modes of a representative and thus are clearly not invariant quantities along the orbit. Unfortunately, we don't know such a convenient criterion of this kind.

However, if we look at the { killing vector of the metric instead of looking at the metric itself}, we can find a practical method, at least for the orbits other than Diff$(S^1)/S^1$, as we will see in what follows.

\subsubsection*{Diffeomorphism invariants associated to Killing vectors}

As we saw in Section 3 with some detail, the orbits that do not contain constant representatives are defined by a vector field  $f\partial_\theta$ fulfilling one of the two requirements:

\begin{itemize}
\item $f$ has an even number of simple zeros, with $\mid{f'}(x_i)\mid=1$ at each zero.
\item $f$ has double zeros and a vanishing third derivative at each zero.
\end{itemize}
Given a central charge $c=t$, such $f$ defines a caodjoint vector $(b_f d\theta^2,c)$  belonging to one of the mentioned orbits through the following relation \cite{Witten88}

\begin{equation} 
b_f=\frac{c}{24}\frac{ff''-\frac{1}{2}(f'^2 -f'^2(x_0))}{f^2}
\end{equation}
being $x_0$ one of the zeros. As we have said, $f'^2(x_0)$ will be $1$ in the first case and zero in the second case. It can be shown that $b_f$ solves (\ref{coadjointaction}) equated to zero and that it is well defined on the zeros of $f$. In other words, the element $b_f$ defined in this way will have $f$ as the generator of the isotropy group (remember that these orbits have a one-parametric isotropy group). This means that the vector field $f\partial_\theta$ will be the killing vector of the corresponding copy of the metric.

The advantage of describing the space of metrics by a function $f$ resides in that there are numbers computed easily from  $f$ which uniquely characterize  the orbit of the associated $b_f$. These numbers were introduced already in Section 3 and here we recall them:

\begin{itemize}
\item $(n,\Delta)$ for the case of simple zeros
\item $n,+/-$ for the case of double zeros
\end{itemize}
where $2n$ is the number of zeros, $\Delta=\lim_{\epsilon\rightarrow{0}}\int_{S^1-V_{\epsilon}}{\frac{1}{f}}$ is the integral defined in (\ref{Delta}), and $+$ and $-$ are invariant signs explained in \cite{Witten88}.

We can extend this to the case of the orbits with a constant representative such that $b_0=-c n^2/48\pi$. For these orbits there are actually three Killing vectors\\ $\{\partial_\theta,\cos(n\theta)\partial_\theta,\sin(n\theta)\partial_\theta\}$, and only the last two have zeros, which are $2n$ simple zeros. Even more, these two $f$ have $\Delta=0$. So, we can say that, coming back to the gravity application, if we take an $f$ of the first kind with $\Delta=0$ and $n=1$ we will know that the corresponding metric will be an AdS$_3$ graviton, even if we are not able to see which is the diffeomorphism that brings the metric to the one given by $\gamma_{++}=\gamma_{--}=-\frac{1}{4}$. If $\Delta\neq{0}$, we can be sure that the metric is not diffeomorphic to $AdS_3$ since in this case the orbit isotropy group is $T_{n,\Delta}$. BTZ black holes are excluded from the analysis since their corresponding $f$ are just constant functions so they have no zeros. This means that if you start with an $f$ with no zeros (not necessarily constant), the orbit must have a constant representative, and it may correspond to a BTZ black hole, but this is as far as we can say, since it could be a negative constant representative. From what we have just seen, given a Killing vector $f\partial_\theta$, it is easier to conclude the absence of a BTZ black hole on the orbit than the existence. On the other hand, the orbit of AdS$_3$ can be identified without difficulty.

\subsection{Extending boundary exotic geometries}

So far we have concentrated in the asymptotic structure of the metrics, without worrying about possible obstructions to extend them away from the boundary, to the region $r<r_0$ (see discussion at the beginning of Section 4). Here we address this issue by taking advantage of a recent work by Li and Lucietti \cite{Escoceses}. In that work they use a particular set of coordinates introduced in \cite{Escocesesprevio} that can be extended\footnote{We thank J. Lucietti for clarifications about \cite{Escoceses}.} beyond the horizon $r_0$ of a black hole, but only until some $\tilde{r}_0<r_0$. We are going to show now that the family of spacetimes with a degenerate Killing horizon considered in \cite{Escoceses} contains examples of the exotic boundary gravitons we have presented before, and thus these gravitons can be extended beyond the horizon.     

The degenerate solutions of \cite{Escoceses} have an expansion for large $r$ that gives 
\begin{equation}
\gamma_{++}=0,\qquad \gamma_{--}=\beta(x^-)=\frac{1}{\alpha^2}\left(1-\frac{\alpha'}{\alpha}\right)
\end{equation}
where $\alpha$ is an arbitrary function which must satisfy\footnote{What we call $\alpha$ is the function $\gamma_1$ in \cite{Escoceses}.}
\begin{equation}
\alpha>0, \qquad \alpha(x^-+2\pi)=\alpha(x^-)
\end{equation}
We see that one of the copies has the zero constant representative and thus is in a Diff$(S^1)/S^1$ orbit and has $\ell M=-J$. It remains to study the other copy. For this, we want to show that there is a function $\alpha$ such that $\gamma_{--}$ is a representative of the desired orbit.     

Let us change variables as $U=\frac{1}{\alpha^2}$ and then
\begin{equation}
\gamma_{--}=U+\frac{1}{2}U'
\end{equation}
The solution to this equation is
\begin{equation}
U=2e^{-2x}\left(\int_0^{x} e^{2\lambda} \gamma_{--}(\lambda) d\lambda +C\right)
\end{equation}
In order for $U$ to be $2\pi$-periodic we demand
\begin{equation}
C=\frac{\int_0^{2\pi} e^{2\lambda} \gamma_{--}(\lambda)  d\lambda }{e^{4\pi}-1}
\end{equation}
The last thing  we need is $U>0$. We explore this for some representatives of the exotic orbits. 
For example, for $\gamma_{--}=\frac{12\pi}{t}b_{1,\Delta}$  as in (\ref{bforTnDelta}), we numerically see that  for $\Delta=100$ then U is always positive. This means that the orbit Diff$(S^1)/T_{1,100}$ has a representative that can be extended beyond $r_0$.   

Now, let us consider the exotic orbit $\tilde{T}_{1,+}$. We recall that this orbit is the only exotic one with the energy bounded from below and also has the property of possessing representatives being arbitrary close to AdS$_3$. This orbit is also in the family of metrics of \cite{Escoceses} since $U>0$ for some value\footnote{The window here is narrow, and we get $U>0$ for $a=0.399$, with $a$ the parameter of the curve in the orbit.} of $a$ in (\ref{Tmonocurve}). In general, to prove that one representative can be extended beyond $r_0$ is almost all we need to prove  to claim that actually all the elements of the corresponding orbit can be extended too. The only thing left is to show that the diffeomorphism that connects any two elements of the orbit remains well defined beyond $r_0$ as we already discussed at the beginning of the section.

\section{Humble digression on the quantization of the orbits}

In this section we give an incomplete and brief summary of the quantization of Virasoro orbits, while trying to present the most relevant points to our understanding.  The mathematical literature on the topic is vast and deserves a much thorough presentation (for example see \cite{Roger}).

\subsection*{Geometric quantization}
A natural approach for quantizing coadjoint orbits is that called geometric quantization \cite{Woodhouse}.
The geometric quantization scheme mainly aims - by means of a geometrical procedure - to assign to certain classical physical system  described by a symplectic manifold $(M, \omega)$, a quantum theory: a Hilbert space $\mathcal{H}$ and Hermitian operators on it. More specifically, this procedure must define an algebra morphism (up to a factor) between the Poisson algebra of classical observables, i.e.  of real functions on $M$, and Hermitian operators on $\mathcal{H}$,  such that the following requirements are met \cite{Geom}:
\begin{list}{}{}
\item i)   $O_{f_1+f_2}=O_{f_1}+O_{f_2}$
\item ii)  $O_{\lambda f}=\lambda O_f,\quad \lambda \in \mathbb{C}$
\item iii) $O_1=\text{Id}_{ \mathcal{H}}$
\item iv)  $[O_{f_1},O_{f_2}]=i\hbar O_{\{f_1,f_2\}}$
\end{list}
where $O_f$ is the self-adjoint operator corresponding to the classical obervable $f$. A fifth requirement is introduced in order to respect the symmetries of the classical phase space
\begin{list}{}{}
\item v) If there is a symmetry group $G$ of symplectomorphisms that act on $(M,\omega)$, then $(\mathcal{H},U)$ should provide an irreducible unitary representation of $G$, where $U:G \rightarrow U(\mathcal{H})$ is a group homomorphism.
\end{list}
This requirements should hold for any quantization with the same spirit as that of the canonical quantization. The particularity of the geometric quantization procedure resides in the geometrical structures introduced in order to define a Hilbert space and the Hermitian operators. We are not going to give a detailed account of the scheme, we just want to mention the key points. First, starting from  a  $2n$-dimensional symplectic manifold $(M,\omega)$, one needs to construct a line bundle $L$ with base $M$, a curvature given by $\omega/2\pi\hbar$, and a compatible\footnote{The necessary and sufficient condition for $\omega/2\pi\hbar$ to be the curvature of the line bundle we are describing, namely with hermitian structure and a compatible connection, is that the cohomology class $[\frac{\omega}{2\pi\hbar}]\in H^2(M,\mathbb{R})$ is integer.} hermitian structure $h$. The pre-Hilbert space is the space of sections of compact support of this line bundle and the inner product is given by 
\begin{equation}\label{innerproduct}
\langle \phi_1,\phi_2\rangle:=\left(\frac{1}{2\pi\hbar}\right)^n\,\int_{M}h(\phi_1,\phi_2)\Lambda_\omega, 
\end{equation}   
where $\phi_{1,2}$ are sections of $L$ and $\Lambda_\omega$ is the Liouville volume form $\Lambda_\omega=(-1)^{\frac{1}{2}n(n-1)}\frac{1}{n!}\omega^n$. To obtain a Hilbert space $\mathcal{H}_\mathcal{P}$ one needs to take the completion of this pre-Hilbert space. This is what is known as pre-quantization and $\mathcal{H}_\mathcal{P}$ is the \textit{prequantum Hilbert space}. 
The self-adjoint operators that satisfy the (first four) conditions stated above are given by
\begin{equation}\label{quantumobservables}
O_f=-i\hbar \nabla_{X_f}+f
\end{equation}    
where $\nabla$ is the connection on $L$ and $X_f$ is the Hamiltonian vector field of $f$.

The next part of the geometric quantization program is  called \textit{polarization}. This is introduced because the pre-quantum Hilbert space and its quantum observables (\ref{quantumobservables}) fail to satisfy the fifth condition regarding irreducibility. The idea is to considerably reduce  the space $\mathcal{H}_\mathcal{P}$ so that it becomes an irreducible representation of the symmetry group $G$. Of course, the implementation of this idea is through a polarization $P$:
\begin{equation}
\Gamma_P(L)=\{\phi \in \Gamma(L) \,| \,\nabla_X\phi=0, \,\forall X \in P  \}
\end{equation}
The Hilbert space $\mathcal{H}$ is then defined as 
$$\mathcal{H}=\Gamma_P(L) \cap \mathcal{H}_\mathcal{P}$$ 
The quantum observables allowed by the polarization are those that respect it:
\begin{equation}
[O_f,\nabla_P]\phi=0,\qquad \forall \phi \in \Gamma_P(L)
\end{equation}  
This condition on the self-adjoint operators translates into a restriction on the space of classical observables. In other words, once a polarization is chosen, not all the classical observables can be mapped to a Hermitian operator. If the symplectic manifold happens to be K\"ahler, then one has a natural polarization called \textit{K\"ahler polarization}, which consists on keeping only the vector fields that belong to one of the eigenspaces of the complex structure of the manifold. 

There is a third step of the geometric quantization program called \textit{metaplectic correction} but since it does not seem to play an important part in the present discussion we omit it.

\subsection*{State of the art of Virasoro orbits quantization: a glance}

The first thing to note is that the coadjoint orbits of Virasoro group are infinite-dimensional manifolds, so the geometric quantization procedure, which was originally thought for finite-dimensional systems, should be adapted. One optimistic fact is that among the Virasoro coadjoint orbits, there are two types which are proved to be K\"ahler \cite{KirillovYurev}, and they are Diff$(S^1)/S^1$ and  Diff$(S^1)/$PSL$_2$. Notably, these are precisely the orbits where AdS$_3$ and the BTZ black holes live. Unfortunately, as far as we know, the geometric quantization of these orbits has not been achieved so far\footnote{Recently a work appeared on positive energy representations of Virasoro group \cite{NeebSalmasian}, which seems to make important improvements towards the geometric quantization of Diff$(S^1)/S^1$.}. The main obstacle, due in part to the infinite-dimensionality, appears to be the difficulty in defining an invariant measure on them\footnote{We would like to thank both J. Mickelsson and D. Pickrell for their comments on this issue and also the former for pointing out \cite{NeebSalmasian}.}.

As an attempt to say something more about the geometric quantization,  in \cite{Witten88} is claimed that if one could achieve the geometric quantization  of  the AdS$_3$ orbit for $c>1$, then its partition function  $\text{Tr}\,q^{\hat L_0}$ would necessarily coincide with that obtained from the Verma module representation :

\begin{equation}\label{AdSpartitionfunction}
    \text{Tr}\,q^{\hat L_0}=\prod_{n=2} \frac{1}{1-q^n}
\end{equation}
This is precisely the partition function of the AdS$_3$ sector used in \cite{MaloneyWitten} to compute, through modular transformations, the contributions to the partition function of the Euclidean black holes. 

The Verma module provides the usual quantization found in the physics literature  by giving a representation of the \textit{Virasoro algebra}  (see for example \cite{Schottenloher}). This module possesses a Hilbert space structure where $i\hat L_n$ and $i\hat Z$ are the quantum versions of the classical functions on phase space $\Phi_{L_n}$ and $\Phi_Z$ respectively. However, what is not available in this quantization is the map from the \textit{Virasoro group} symmetries to a unitary irreducible representations as demanded by the fifth requirement above. Nevertheless, the algebra representation through the Verma module  can be integrated to  a unitary representation of Virasoro group according to \cite{GW85}. Even more, these unitary representations are contained in the classification  of \cite{NeebSalmasian} which actually makes contact with the geometric quantization of the Diff$(S^1)/S^1$ orbit. In a future work we want to explore what this can tell about the quantization of the orbits of AdS$_3$ gravity.

\section*{Acknowledgments}

We are grateful to G. Barnich, M. Farinati, J. Lucietti, L. Lombardi, J. Mickelsson, D. Pickrell and specially to G. Giribet and  C. Roger for enlightening discussions. A. G. is thankful for the hospitality of the Physics Department of the University of Buenos Aires. This work was partially supported by grants PIP and PICT from CONICET and ANPCyT.

\end{document}